\documentclass[12pt]{article}
\usepackage{epsfig,latexsym,amsmath,amssymb,bm,cite}
\usepackage{graphicx}
\usepackage{color}
\newcommand{\be}{\begin{equation}}
\newcommand{\ee}{\end{equation}}
\newcommand{\bea}{\begin{eqnarray}}
\newcommand{\eea}{\end{eqnarray}}
\newcommand{\bem}{\begin{multline}}
\newcommand{\eem}{\end{multline}}
\newcommand{\beg}{\begin{gather}}
\newcommand{\eeg}{\end{gather}}

\newcommand{\stackeven}[2]{{{}_{\displaystyle{#1}}\atop\displaystyle{#2}}}
\newcommand{\lsim}{\stackeven{<}{\sim}}

\def\eq#1{{Eq.~(\ref{#1})}}
\def\fig#1{{Fig.~\ref{#1}}}
\newcommand{\ben}{\begin{eqnarray*}}
\newcommand{\een}{\end{eqnarray*}}
\newcommand{\un}[1]{\underline{#1}}

\begin{document}
\title{{\bf Early Time Dynamics in Heavy Ion Collisions \\[.5cm] 
from AdS/CFT Correspondence
\\[1.5cm] }}
\author{
{\bf Yuri V.\ Kovchegov\thanks{e-mail: yuri@mps.ohio-state.edu} \ 
and Anastasios Taliotis\thanks{e-mail: taliotis.1@osu.edu}}
\\[1cm] {\it\small Department of Physics, The Ohio State University}\\ 
{\it\small Columbus, OH 43210,USA}\\[5mm]}

\date{May 2007}

\maketitle

\thispagestyle{empty}

\begin{abstract}
  We study the matter produced in heavy ion collisions assuming that
  this matter is strongly interacting and employing AdS/CFT
  correspondence to investigate its dynamics. At late proper times
  $\tau$ we show that Bjorken hydrodynamics solution, obtained
  recently by Janik and Peschanski using gauge-gravity duality
  \cite{Janik:2005zt}, can be singled out by simply requiring that the
  metric tensor is a real and single-valued function of the
  coordinates everywhere in the bulk, without imposing any constraints
  on the curvature invariant.  At early proper times we use similar
  strategy to show that the energy density $\epsilon$ approaches a
  constant as $\tau \rightarrow 0$. We therefore demonstrate that the
  strong coupling dynamics incorporates the isotropization transition
  in heavy ion collisions.  By matching our early-time regime with the
  late-time one of Janik and Peschanski we estimate the isotropization
  time at RHIC to be approximately $\tau_{\text{iso}} \approx
  0.3$~fm/c, in good agreement with results of hydrodynamic
  simulations.
\end{abstract}

\thispagestyle{empty}

\newpage

\setcounter{page}{1}

%%%%%%%%%%%%%%%%%%%%%%%%%%%%%%%%%%%%%%%%%%%%%%%%%%%%%%%%%%%%%%%%%%%%%%%%%%%%%%%%%

\section{Introduction}

The problem of thermalization of the system produced in heavy ion
collisions is crucial for our understanding of strong interactions.
While in recent years there has been some significant progress in
understanding the initial (pre-thermalization) stages of heavy ion
collisions in the framework of the Color Glass Condensate (CGC)
\cite{Blaizot:1987nc,McLerran:1993ni,McLerran:1993ka,McLerran:1994vd,Kovchegov:1996ty,Kovchegov:1997pc,Kovner:1995ja,Kovchegov:1997ke,Krasnitz:1998ns,Krasnitz:1999wc,Krasnitz:2003jw,Kovchegov:2000hz,Lappi:2003bi,Kharzeev:2000ph,Kharzeev:2001yq,Kharzeev:2002pc,Kharzeev:2004yx,Albacete:2003iq}
(for a review of CGC see
\cite{Iancu:2003xm,Weigert:2005us,Jalilian-Marian:2005jf}), a complete
theoretical understanding of the subsequent isotropization and
thermalization of the produced medium continues to evade us. The
success of hydrodynamics simulation in describing the data generated
at RHIC is quite impressive: however, it requires a very early
starting proper time for hydrodynamics, $\tau \lsim 0.5$~fm/c
\cite{Kolb:2000sd,Kolb:2000fh,Huovinen:2001cy,Kolb:2001qz,Heinz:2001xi,Teaney:1999gr,Teaney:2000cw,Teaney:2001av}.
The existing theoretical models based on the weak-coupling dynamics so
far have not been successful in reproducing such an early
thermalization time
\cite{Baier:2000sb,Mrowczynski:1988dz,Mrowczynski:1993qm,Arnold:2003rq,Arnold:2004ti,Rebhan:2004ur,Romatschke:2006nk}.
Moreover, as was suggested by one of the authors
\cite{Kovchegov:2005ss,Kovchegov:2005kn,Kovchegov:2005az}, the
perturbative weak-coupling dynamics, describable by Feynman diagrams,
may in principle be unable to generate isotropization, and, therefore,
thermalization in high energy heavy ion collisions. It was suggested
in \cite{Kovchegov:2005ss,Kovchegov:2005kn,Kovchegov:2005az} that the
only way isotropization and thermalization can be achieved in heavy
ion collisions is due to strong-coupling effects.

Another more phenomenological argument indicating that strong coupling
dynamics could be important for heavy ion collisions at RHIC is the
fact that hydrodynamic description of the collisions
\cite{Kolb:2000sd,Kolb:2000fh,Huovinen:2001cy,Kolb:2001qz,Heinz:2001xi,Teaney:1999gr,Teaney:2000cw,Teaney:2001av}
has to have a very low shear viscosity \cite{Teaney:2003kp}: the
quark-gluon plasma produced at RHIC is almost a perfect fluid. Since
the perturbative shear viscosity scales as $\eta \sim 1/g^4$
\cite{Arnold:2000dr} one concludes that low viscosity requires large
coupling. Such conclusion was confirmed by explicit calculations
employing gauge-gravity duality in
\cite{Policastro:2001yc,Son:2002sd,Policastro:2002se,Kovtun:2003wp,Kovtun:2004de}.

Unfortunately, the strong coupling regime of QCD is not under
theoretical control. However, for some gauge theories, like ${\cal N}
=4$ super Yang-Mills theory, the strong coupling regime can be studied
using AdS/CFT correspondence
\cite{Maldacena:1997re,Gubser:1998bc,Witten:1998qj,Aharony:1999ti}.
To understand the strong coupling dynamics of the matter produced in
heavy ion collisions one has to model QCD with ${\cal N} =4$ super
Yang-Mills theory and use AdS/CFT correspondence to study the
strong-coupling dynamics of such theory. In recent years there have
been many applications of AdS/CFT correspondence to heavy ion
collisions
\cite{Janik:2005zt,Janik:2006gp,Janik:2006ft,Nakamura:2006ih,Shuryak:2005ia,Lin:2006rf,Lublinsky:2007mm,Nastase:2005rp,Bak:2006dn,Kajantie:2006hv,Kajantie:2006ya,Kajantie:2007bn}.

The goal of this paper is to attempt to understand the onset of
isotropization and thermalization in the AdS/CFT framework. We will
consider an idealized Bjorken picture of the collision
\cite{Bjorken:1982qr}, in which nuclei have an infinite extent in the
transverse plane and the distribution of the produced matter is
rapidity-independent: the energy and pressure components of the
energy-momentum tensor in the local rest frame would then depend only
on the proper time $\tau$.  We would like to emphasize that
isotropization and thermalization of the produced medium do not
necessarily happen at the same time: isotropization is necessary for
thermalization, but not vice versa. As was argued in
\cite{Lappi:2003bi}, in the CGC framework the energy density
$\epsilon$ of the matter produced in heavy ion collisions scales as a
power of a logarithm of the proper time $\tau$, as shown in \eq{ecgc}
below. Therefore, the produced matter is anisotropic: its longitudinal
pressure component is in fact negative. Isotropization for such matter
would imply generation of positive longitudinal pressure, comparable
to the transverse pressure components. Note that for isotropization to
take place it does not appear necessary for the system to have a
thermal distribution: hence thermalization may take place after
isotropization, and not necessarily at the same time. Below we will
study the onset of isotropization, since it is much easier to infer
from the components of the energy-momentum tensor.

In a pioneering paper \cite{Janik:2005zt} Janik and Peschanski showed
that for the strongly-coupled medium produced in heavy ion collisions
and described by gauge-gravity duality the late time asymptotics can
only lead to Bjorken hydrodynamics \cite{Bjorken:1982qr} with the
energy density scaling with proper time as $\epsilon \sim
\tau^{-4/3}$. Since then a number of further investigations have been
carried out \cite{Nakamura:2006ih,Janik:2006ft,Heller:2007qt}, mostly
concentrating on the subleading at late times viscous corrections to
the ideal hydrodynamic behavior of \cite{Janik:2005zt}. In this work
we will take a different route and concentrate on the early time
behavior of the strongly coupled medium produced in heavy ion
collisions.  The hope is to verify if such medium is anisotropic at
early proper times: if that proves to be the case then the onset of
isotropization would necessarily be a part of strongly-coupled
medium's dynamics at intermediate proper times. The idea of studying
the matter produced in nuclear collisions at small proper times has
been originally suggested in the CGC framework in
\cite{Fries:2006pv,Lappi:2003bi}.  Here we will apply the same concept
to the strong-coupling dynamics employing the AdS/CFT correspondence.

The paper is structured as follows. We begin in Sect. \ref{general} by
setting up the formalism of holographic renormalization in the context
of heavy ion collisions with Bjorken geometry. We then proceed in
Sect. \ref{power} by solving Einstein equations (\ref{eom1}) in the
AdS space assuming that the four-dimensional energy density scales as
a power of proper time, as given by \eq{etau}. In Sect. \ref{iter} we
will demonstrate the first few steps of the perturbative solution
expanding the metric in powers of the Fefferman-Graham coordinate $z$
\cite{F-G}. In Sect \ref{Bjorken} we proceed by re-analyzing the late
proper time dynamics: we show that in order to pick the correct
Bjorken hydrodynamic behavior of the solution of Einstein equations
(\ref{eom1}) one only has to impose the condition that the resulting
metric is a real single-valued function of the coordinates everywhere
in the bulk.  Hence we demonstrate that the condition of
non-singularity of the curvature invariant employed in
\cite{Janik:2005zt} is not needed to constrain the dynamics in four
dimensions from AdS/CFT. We then apply the same approach to study the
early time dynamics in Sect.  \ref{earlytime}: we solve Einstein
equations (\ref{eom1}) at small-$\tau$ and, after imposing the
condition for the metric to be real and single-valued, we derive that
energy density can only go to a constant value as $\tau \rightarrow 0$
(see \eq{earlye}).

This is our main result: as the system having constant energy density
in heavy ion collisions also has negative longitudinal pressure, we
conclude that at very early times the strongly interacting system
described by gauge-gravity duality is very anisotropic. At the same
time we know that at late proper times it becomes isotropic evolving
according to Bjorken hydrodynamics. We thus demonstrate that the
isotropization transition does take place at intermediate times for
the evolution of this strongly-coupled system! Moreover, while indeed
the early-time energy density in CGC [\eq{ecgc}] and in the
strongly-coupled medium [\eq{earlye}] behave somewhat differently, the
two initial-time behaviors are not entirely unlike each other: in both
cases the initial longitudinal pressure is negative, equal to the
negative of the energy density, while the transverse pressures are
equal to the energy density [see \eq{earlypp3}]. We therefore conclude
that the behavior of the strongly-coupled system studied here may be
relevant for real-life heavy ion collisions, where the initial stages
of the collisions are successfully described by the CGC. 

We verify our result of constant energy density at early times in
Sect. \ref{log}, where, inspired by CGC \cite{Lappi:2003bi}, we look
for possible logarithmic scaling of the energy density at early time,
but find none.  The energy density remains constant at small $\tau$,
without any logarithmic divergence.

In Sect. \ref{iso_time} we match the early time dynamics considered in
this work onto the late-time Bjorken hydrodynamic solution found in
\cite{Janik:2005zt} to estimate the isotropization time (see
\fig{iso}). The result is given by \eq{tiso}. If we apply this formula
to central $Au+Au$ collisions at RHIC at $\sqrt{s} = 200$~GeV we
obtain the isotropization time of $\tau_{\text{iso}} \approx
0.3$~fm/c, in good agreement with results of hydrodynamic simulations.

We conclude in Sect. \ref{conc} by restating our main results and by
summarizing our knowledge of energy density as a function of proper
time for the strongly-coupled system at hand in \fig{escaling}.

%%%%%%%%%%%%%%%%%%%%%%%%%%%%%%%%%%%%%%%%%%%%%%%%%%%%%%%%%%%%%%%%%%%%%%%%%%%%%%%%%

\section{Holographic renormalization: general setup}
\label{general}

Let us consider a head-on collision of two very large identical
nuclei. For simplicity we assume that the distribution of matter
produced in such collisions is independent of (space-time) rapidity
$\eta$. Since the nuclei are identical the matter distribution should
also respect $x_+ \leftrightarrow x_-$ ($\eta \leftrightarrow -\eta$)
symmetry. Finally, as the nuclei are very large, we assume that they
have infinite extent in the transverse directions. Therefore the
energy-momentum tensor of the produced matter is independent of the
transverse coordinates ${\un x} = (x_1, x_2)$. This is indeed the same
approximation as employed by Bjorken in \cite{Bjorken:1982qr}.

Following Janik and Peschanski \cite{Janik:2005zt} we write the metric
in AdS$_5$ space for such a system using Fefferman--Graham coordinates
\cite{F-G} as
\begin{align}\label{met_gen}
  d s^2 \, = \, \frac{1}{z^2} \, \left[ - A (\tau, z) \, d \tau^2 +
    \tau^2 \, B (\tau, z) \, d \eta^2 + C (\tau, z) \, d x_\perp^2 + d
    z^2\right]
\end{align}
where $\tau = \sqrt{2 \, x_+ \, x_-}$ is the proper time in four
dimensions, $\eta = (1/2) \ln (x_+/x_-)$ is the space-time rapidity,
$x_\pm = (x_0 \pm x_3)/\sqrt{2}$, $d x_\perp^2 = d x_1^2 + d x_2^2$
and $z$ is the coordinate in the $5$th dimension. Just like in
\cite{Janik:2005zt} we write
\begin{align}\label{ABC}
  A (\tau, z) \, = \, e^{a (\tau, z)}, \ \ \ B (\tau, z) \, = \, e^{b
    (\tau, z)}, \ \ \ C (\tau, z) \, = \, e^{c (\tau, z)},
\end{align}
where $a (\tau, z)$, $b (\tau, z)$ and $c (\tau, z)$ are some unknown
functions, to be determined from the solution of Einstein equations
with a negative cosmological constant $\Lambda = -6$ in AdS$_5$ space
\begin{align}\label{eom1}
  R_{\mu\nu} - \frac{1}{2} \, g_{\mu\nu} \, R - 6 \, g_{\mu\nu} = 0.
\end{align}
Here $R_{\mu\nu}$ is the Ricci tensor and $R$ is the scalar curvature.
Contracting \eq{eom1} with $g^{\mu\nu}$ yields
\begin{align}
  R = - 20
\end{align}
which, when inserted back into \eq{eom1} gives
\begin{align}\label{eom}
  R_{\mu\nu} + 4 \, g_{\mu\nu} = 0. 
\end{align}

Finding an exact general solution of \eq{eom} with the metric ansatz
(\ref{met_gen}) appears to be a daunting task. 
%Since here we are
%interested in early times (small $\tau$) dynamics, 
Instead, we begin by exploring the solution using the technique of
holographic renormalization \cite{deHaro:2000xn}. The general
prescription for a metric of the form
\begin{align}
  d s^2 \, = \, \frac{1}{z^2} \, \left[ {\tilde g}_{\mu\nu} (x, z) \,
    d x^\mu \, d x^\nu + d z^2\right]
\end{align}
is to expand it near the $z=0$ boundary in a series
\cite{deHaro:2000xn}
\begin{align}\label{hr_series}
  {\tilde g}_{\mu\nu} (x, z) \, = \, {\tilde g}_{\mu\nu}^{(0)} (x) +
  z^2 \, {\tilde g}_{\mu\nu}^{(2)} (x) + z^4 \, {\tilde
    g}_{\mu\nu}^{(4)} (x) + \ldots
\end{align}
and solve Einstein equations (\ref{eom}) order-by-order in $z$. 

In our case we know the metric on the $z=0$ boundary: it is simply the
flat Minkowski metric, such that ${\tilde g}_{\mu\nu}^{(0)} (x) =
\eta_{\mu\nu}$. Then one can show that ${\tilde g}_{\mu\nu}^{(2)} (x)
= 0$ \cite{deHaro:2000xn}. Finally, the fourth order term in
\eq{hr_series} is related to the expectation value of the
energy-momentum tensor $T_{\mu\nu}$ in our four dimensions
\cite{deHaro:2000xn,Janik:2006ft}
\begin{align}\label{g4}
  \langle T_{\mu\nu} \rangle \, = \, \frac{N_c^2}{2 \, \pi^2} \
  {\tilde g}_{\mu\nu}^{(4)} (x)
\end{align}
with $N_c$ the number of colors. As can be easily shown
\cite{Kovchegov:2005ss}, the most general energy-momentum tensor in
the boost-invariant geometry of the collision of two very large
nuclei can be written (in the co-moving frame) as 
\begin{eqnarray}\label{tmn_gen}
   \langle T^{\mu\nu} \rangle &=&  
 \left( 
\begin{matrix}
  \epsilon (\tau) & 0 & 0 & 0 \cr
  0 & p (\tau) & 0 & 0 \cr
  0 & 0 & p (\tau) & 0  \cr
  0 & 0 & 0  & p_3 (\tau) \cr 
\end{matrix}
\right)
\end{eqnarray}
in the $t, x_1, x_2, x_3$ coordinate system. Here $\epsilon (\tau)$ is
the energy density, $p (\tau)$ is the transverse pressure component,
and $p_3 (\tau)$ is the longitudinal pressure component. (The latter
two quantities do not have to be equal in general: they are equal only
in the case of ideal hydrodynamics.) 
%Comparing Eqs. (\ref{tmn_gen}) and
%(\ref{g4}) we conclude that
%\begin{align}
%  \epsilon (\tau) \, = \, - \frac{N_c^2}{2 \, \pi^2} \ {\tilde
%    g}_{00}^{(4)} (\tau), \ \ \ p (\tau) \, = \, \frac{N_c^2}{2 \,
%    \pi^2} \ {\tilde g}_{11}^{(4)} (\tau) \, = \, \frac{N_c^2}{2 \,
%    \pi^2} \ {\tilde g}_{22}^{(4)} (\tau), \ \ \ p_3 (\tau) \, = \,
%  \frac{N_c^2}{2 \, \pi^2} \ {\tilde g}_{33}^{(4)} (\tau).
%\end{align}

In \cite{Janik:2005zt} it was elegantly shown that requiring
non-negativity of the energy density $\epsilon (\tau)$ in all frames
leads to the following conditions on its dependence on the proper time
$\tau$
\begin{align}\label{conds}
  \epsilon' (\tau) \, \le \, 0, \hspace*{1cm} \tau \, \epsilon' (\tau)
  \, \ge \, - 4 \, \epsilon (\tau).
\end{align}
We will use this conditions later to restrict energy density. 

As will be manifest shortly, Einstein equations lead to the following
conditions for the energy-momentum tensor. First of all they require
energy-momentum conservation in four dimensions
\begin{align}
  \partial_\mu \, \langle T^{\mu\nu} \rangle \, = \, 0
\end{align}
which, when applied to \eq{tmn_gen}, gives
\begin{align}\label{hydroeq}
  \frac{d \epsilon}{d \tau} \, = \, - \frac{\epsilon + p_3}{\tau}.
\end{align}
(Here and below we will sometimes suppress the arguments of
$\epsilon$, $p$ and $p_3$.)  Secondly they require the energy-momentum
tensor to be traceless
\begin{align}
  \langle T_\mu^{\mu} \rangle \, = \, 0, 
\end{align}
which, for \eq{tmn_gen} leads to
\begin{align}\label{trless}
  \epsilon \, = \, 2 \, p + p_3.
\end{align}

Finally, to apply the expansion of \eq{hr_series} with ${\tilde
  g}_{\mu\nu}^{(0)} (x) = \eta_{\mu\nu}$ and ${\tilde
  g}_{\mu\nu}^{(2)} (x) = 0$ to the metric in \eq{met_gen} we write
\cite{Janik:2005zt}
\begin{align}\label{abc}
  a (\tau, z) \, = \, \sum\limits_{n=0}^\infty a_n (\tau) \, z^{4 + 2
    n}, \ \ \ b (\tau, z) \, = \, \sum\limits_{n=0}^\infty b_n (\tau)
  \, z^{4 + 2 n}, \ \ \ c (\tau, z) \, = \, \sum\limits_{n=0}^\infty
  c_n (\tau) \, z^{4 + 2 n}.
\end{align}
Comparing these to Eqs. (\ref{g4}) and (\ref{tmn_gen}) yields
\begin{align}\label{epp3}
  \epsilon \, = \, - \frac{N_c^2}{2 \, \pi^2} \ a_0 (\tau), \ \ \ p_3
  \, = \, \frac{N_c^2}{2 \, \pi^2} \ b_0 (\tau), \ \ \ p \, = \,
  \frac{N_c^2}{2 \, \pi^2} \ c_0 (\tau).
\end{align}

Now we are ready to implement the holographic renormalization using a
particular ansatz for the energy density.

%%%%%%%%%%%%%%%%%%%%%%%%%%%%%%%%%%%%%%%%%%%%%%%%%%%%%%%%%%%%%%%%%%%%%%%%%%%%%%%%%

\section{Power-law scaling of energy density with proper time}
\label{power}

%%%%%%%%%%%%%%%%%%%%%%%%%%%%%%%%%%%%%%%%%%%%%%%%%%%%%%%%%%%%%%%%%%%%%%%%%%%%%%%%%

\subsection{First steps: iterative solution}
\label{iter}

Let us first assume that the energy density scales as
\begin{align}\label{etau}
  \epsilon (\tau) \, \sim \, \tau^\Delta. 
\end{align}
This is the simplest assumption. We know that it works at late times
$\tau \gg 1$ (in units of $\sqrt{G_N}$ with $G_N$ the Newton's
constant in AdS$_5$) \cite{Janik:2005zt}. In a little while we will
try to apply it to the early-time dynamics ($\tau \ll 1$) and if it
would not work we will be forced to look for a more sophisticated one.
At the moment we are not going to make any assumptions or
approximations specific to early or late times.

We will begin with the scaling of \eq{etau} and try to build a
solution of Einstein equations in the bulk using holographic
renormalization. From the dynamics in four dimensions we know that if
$\epsilon \sim \tau^\Delta$ then $p \sim \tau^\Delta$ and $p_3 \sim
\tau^\Delta$.  Using \eq{epp3} we write
\begin{align}\label{abc0}
  a_0 (\tau) \, = \, a_0 \, \tau^\Delta , \ \ \ b_0 (\tau) \, = \, b_0
  \, \tau^\Delta , \ \ \ c_0 (\tau) \, = \, c_0 \, \tau^\Delta
\end{align}
with $a_0$, $b_0$ and $c_0$ some constants. Substituting \eq{abc0}
into \eq{abc}, and using the expansion from the latter in \eq{eom} we
solve Einstein equations to the lowest non-trivial order in $z$. This gives
\begin{subequations}\label{abc00}
\begin{align}
  a_0 (\tau) \, & = \, a_0 \, \tau^\Delta \\
  b_0 (\tau) \, & = \, a_0 \, (\Delta + 1) \, \tau^\Delta \\
  c_0 (\tau) \, & = \, - a_0 \, \frac{\Delta + 2}{2} \, \tau^\Delta.
\end{align}
\end{subequations}
Indeed Eqs. (\ref{abc00}) could have been obtained from Eqs.
(\ref{hydroeq}) and (\ref{trless}) directly from the four-dimensional
considerations of dynamics without trace anomaly. Here we obtain them
from the AdS/CFT correspondence.  

Substituting Eqs. (\ref{abc00}) back into \eq{abc}, and again solving
Einstein equations (\ref{eom}) at the lowest non-trivial order in $z$
(which is now different from the previous step) yields
\begin{subequations}\label{abc1}
\begin{align}
  a_1 (\tau) \, & = \, a_0 \, \frac{\Delta \, (\Delta + 2)}{12} \,
  \tau^{\Delta - 2} \\
  b_1 (\tau) \, & = \, a_0 \, \frac{\Delta \, (\Delta + 2) \, (\Delta
    - 1)}{12} \,  \tau^{\Delta - 2} \\
  c_1 (\tau) \, & = \, - a_0 \, \frac{\Delta^2 \, (\Delta + 2)}{24} \,
  \tau^{\Delta -2}.
\end{align}
\end{subequations}

For what we intend to do next it is instructive to iterate the
procedure several more times. Here we will show only the result of the
next iteration:
\begin{subequations}\label{abc2}
\begin{align}
  a_2 (\tau) \, & = \, - \frac{1}{384} \left[ a_0 \, \tau^{\Delta - 4}
    \, (4 \, \Delta^2 - \Delta^4 ) + a_0^2 \, \tau^{2 \, \Delta} \, 8
    \,  (8 + 8 \, \Delta + 3 \, \Delta^2) \right] \\
  b_2 (\tau) \, & = \, - \frac{1}{384} \left[ a_0 \, \tau^{\Delta - 4}
    \, (-12 \, \Delta^2 + 4 \,\Delta^3 + 3 \Delta^4 - \Delta^5 ) +
    a_0^2 \, \tau^{2 \, \Delta} \, 8
    \,  (8 + 8 \, \Delta + 3 \, \Delta^2) \right] \\
  c_2 (\tau) \, & = \, - \frac{1}{768} \left[ a_0 \, \tau^{\Delta - 4}
    \, (8 \, \Delta^2 - 4 \,\Delta^3 - 2 \Delta^4 + \Delta^5 ) +
    a_0^2 \, \tau^{2 \, \Delta} \, 16
    \,  (8 + 8 \, \Delta + 3 \, \Delta^2) \right].
\end{align}
\end{subequations}

As follows from the conditions in \eq{conds} \cite{Janik:2005zt},
positivity of energy density $\epsilon$ combined with the power-law
ansatz of \eq{etau} requires that 
\begin{align}
  -4 \le \Delta \le 0.
\end{align}
Let us assume that $\Delta > -4$.  This is a safe assumption: at late
times the smallest possible value of $\Delta$ is $\Delta = -4/3$
\cite{Bjorken:1982qr}.  At late times we therefore have $-4 < \Delta
\le 0$.

At early times $\Delta \ge -1$, the total energy of the matter
produced in a collision approximately scales as $E \sim \epsilon \,
\tau$ for small $\tau$. For $\Delta < -1$ the total energy $E$ would
be infinite in the $\tau \rightarrow 0$ limit, which is impossible.
Hence the physically relevant values of $\Delta$ at early times are
constrained to $-1 \le \Delta \le 0$.

%%%%%%%%%%%%%%%%%%%%%%%%%%%%%%%%%%%%%%%%%%%%%%%%%%%%%%%%%%%%%%%%%%%%%%%%%%%%%%%%%

\subsection{Re-deriving Bjorken hydrodynamics}
\label{Bjorken}

Let us first re-visit the dynamics at late times, which has already
been studied in \cite{Janik:2005zt}. If $\Delta > -4$, then the
$\tau^{2 \Delta}$ term dominates over $\tau^{\Delta -4}$ term in
\eq{abc2}, such that Eqs.  (\ref{abc00}), (\ref{abc1}) and
(\ref{abc2}) combined with \eq{abc} give us the following type of a
series for $a (\tau, z)$ in the large-$\tau$ limit
\begin{align}
  a (\tau, z) \, = \, \# \, z^4 \, \tau^\Delta + \#' \, z^8 \, \tau^{2
    \, \Delta} + \ldots 
\end{align}
with $\#$ and $\#'$ some $\Delta$-dependent constants. (The series for
$b(\tau, z)$ and $c(\tau, z)$ are similar.) One can see that $a (\tau,
z)$ becomes a function of a single scaling variable $v \propto z \,
\tau^{\Delta /4}$, as was observed and utilized in
\cite{Janik:2005zt}.

Following Janik and Peschanski \cite{Janik:2005zt} we define the
scaling variable as
\begin{align}\label{v}
  v \, = \, (-a_0)^{1/4} \, z \, \tau^{\Delta /4}.
\end{align}
Note that combining Eqs. (\ref{epp3}) and (\ref{abc00}a) we get
\begin{align}\label{escal}
  \epsilon (\tau) \, = \, - \frac{N_c^2}{2 \, \pi^2} \ a_0 \,
  \tau^\Delta
\end{align}
such that positivity of the energy density requires that $a_0 < 0$ and
the power of $-a_0$ in \eq{v} is well-defined. Assuming that $a (\tau,
z) \, = \, a (v)$, $b (\tau, z) \, = \, b (v)$, and $c (\tau, z) \, =
\, c (v)$ the authors of \cite{Janik:2005zt} solved the Einstein
equations (\ref{eom}) for the metric of Eqs. (\ref{met_gen}) and
(\ref{ABC}), keeping $v$ constant and taking $\tau \rightarrow \infty$
limit. This led to the following solution obtained in
\cite{Janik:2005zt}
\begin{subequations}\label{JPsol}
\begin{align}
  a(v) \, & = \, \frac{1}{2} \, \left( 1 - \frac{1}{D} \right) \, \ln (1
  + D \, v^4) + \frac{1}{2} \, \left( 1 + \frac{1}{D} \right) \, \ln
  (1 - D \, v^4) \\
  b(v) \, & = \, \frac{1}{2} \, \left( 1 - \frac{\Delta + 1}{D} \right)
  \, \ln (1 + D \, v^4) + \frac{1}{2} \, \left( 1 + \frac{\Delta +
      1}{D} \right) \, \ln (1 - D \, v^4) \\
  c(v) \, & = \, \frac{1}{2} \, \left( 1 + \frac{\Delta + 2}{2 \, D} \right)
  \, \ln (1 + D \, v^4) + \frac{1}{2} \, \left( 1 - \frac{\Delta +
      2}{2 \, D} \right) \, \ln (1 - D \, v^4),
\end{align}
\end{subequations}
where
\begin{align}\label{D}
  D \, = \, \sqrt{\frac{3 \Delta^2 + 8 \Delta + 8}{24}}.
\end{align}

To determine the allowed values of the power $\Delta$ let us note that
$\ln (1 - D \, v^4)$ has a branch cut along the real axis for $1 - D
\, v^4 \le 0$.  Generally speaking this means that $A (\tau, z)$, $B
(\tau, z)$ and $C (\tau, z)$ will be complex-valued for $1 - D \, v^4
< 0$. Since the $\tau \rightarrow \infty$ limit taken in arriving at
\eq{JPsol} was done keeping $v$ fixed there is no constraint
prohibiting $v > D^{-1/4}$ and preventing the metric from becoming
complex. The only way for the metric to be real and single-valued for
all $v$ is if the coefficients in front of $\ln (1 - D \, v^4)$ in
\eq{JPsol} are {\sl integers}.  We therefore require that
\begin{subequations}\label{integers}
\begin{align}
  \frac{1}{2} \, \left( 1 + \frac{1}{D} \right) \, = \, n \\
  \frac{1}{2} \, \left( 1 + \frac{\Delta + 1}{D} \right) \, = \, m \\
  \frac{1}{2} \, \left( 1 - \frac{\Delta + 2}{2 \, D} \right) \, = \,
  l
\end{align}
\end{subequations}
with $n, m, l$ some integer numbers. After some straightforward
algebra one derives
\begin{align}
  n + m \, = \, 2 \, (1-l).
\end{align}
Therefore $n + m$ is an even number. Then $n-m$ is also an even
number. We then write using Eq. (\ref{integers}b)
\begin{align}
  - \frac{\Delta}{2 \, D} \, = \, n - m \, = \, 2 \, k
\end{align}
with $k$ some other integer. Solving 
\begin{align}\label{qeq}
  \Delta \, = \, - 4 \, D \, k
\end{align}
for $\Delta$ with the help of \eq{D} yields
\begin{align}\label{dsol}
  \Delta \, = \, - \frac{4}{3} \, \frac{k}{2 \, k^2 - 1} \, \left[ 2
    \, k \pm \sqrt{3 - 2 \, k^2} \right].
\end{align}
As $\Delta \le 0$ and $D \ge 0$ we conclude from \eq{qeq} that $k \ge
0$. For the solutions in \eq{dsol} to be real, the only allowed
non-negative values of $k$ are $k=0, 1$.

If $k=0$ then $\Delta = 0$ and $D = 1/\sqrt{3}$. However, then Eq.
(\ref{integers}a) will not be satisfied as $n$ will not be integer.

If $k=1$ the allowed values of $\Delta$ are $\Delta = -4$ and $\Delta
= -\frac{4}{3}$. However, as we concluded from, say, \eq{abc2}, for
$\Delta = -4$ the solution of \eq{JPsol} is no longer dominant, as
terms subleading for larger values of $\Delta$ become comparable to
it. On top of that $\Delta = -4$ is not a viable physical value of the
power. 

We are then left only with $\Delta = -\frac{4}{3}$, which, when used
in \eq{escal}, gives us
\begin{align}\label{ebj}
  \epsilon (\tau) \, = \, - \frac{N_c^2}{2 \, \pi^2} \ a_0 \,
  \frac{1}{\tau^{4/3}} \, \propto \frac{1}{\tau^{4/3}}
\end{align}
characteristic of ideal Bjorken hydrodynamics \cite{Bjorken:1982qr}.
We have thus obtained the perfect fluid behavior for the matter
produced in heavy ion collisions without imposing a constraint of the
absence of singularities in the curvature invariant ${\cal R} =
R_{\mu\nu\rho\sigma} \, R^{\mu\nu\rho\sigma}$, like it was done in the
original derivation of \cite{Janik:2005zt}. The only necessary
requirement was that the metric tensor is a real and single-valued
function of $v$ for all values of the scaling variable $v$.

%%%%%%%%%%%%%%%%%%%%%%%%%%%%%%%%%%%%%%%%%%%%%%%%%%%%%%%%%%%%%%%%%%%%%%%%%%%%%%%%%

\subsection{Solution at early times: scaling ansatz}
\label{earlytime}

We will now proceed towards the main goal of this work, which is
studying the early-time asymptotics, $\tau \ll 1$. For small $\tau$
the dominant part of the holographic renormalization series studied in
Sect. \ref{iter} is
\begin{align}\label{aser}
  a (\tau, z) \, & = \, \# \, z^4 \, \tau^\Delta + \#' \, z^6 \,
  \tau^{\Delta - 2} + \#'' \, z^8 \, \tau^{\Delta - 4} + \ldots \notag
  \\ & = z^4 \, \tau^\Delta \, \left( \# + \#' \, \frac{z^2}{\tau^2} +
    \#'' \, \frac{z^4}{\tau^4} + \ldots \right)
\end{align}
with $\#$, $\#'$ and $\#''$ some (different) $\Delta$-dependent
constants. (The series for $b(\tau, z)$ and for $c(\tau, z)$ have the
same structure.) Unfortunately, no single scaling variable can
describe the $\tau$ and $z$ dependence now. Nevertheless, as one can
see from \eq{aser}, the series is in $z/\tau$. Hence, defining a new
scaling variable
\begin{align}\label{u}
  u \equiv \frac{z}{\tau}
\end{align}
we rewrite \eq{aser} as
\begin{align}\label{ascal}
  a (\tau, u) \, = \, \tau^{\Delta + 4} \, u^4 \, \left( \, \# + \#'
    \, u^2 + \#'' \, u^4 + \ldots \, \right).
\end{align}
Therefore, a general ansatz for $a (\tau, u)$ at early times is
\begin{align}\label{aa}
   a (\tau, u) \, = \, \tau^{\Delta + 4} \, \alpha (u)
\end{align}
where $\alpha (u)$ is some unknown function. Similarly we write
\begin{align}\label{bc}
  b (\tau, u) \, = \, \tau^{\Delta + 4} \, \beta (u), \ \ \ c (\tau,
  u) \, = \, \tau^{\Delta + 4} \, \gamma (u).
\end{align}
The initial conditions for $\alpha (u)$, $\beta (u)$ and $\gamma (u)$
can be determined from Eqs. (\ref{abc00}) yielding
\begin{align}\label{init}
  \alpha (u) \, = \, a_0 \, u^4, \ \ \ \beta (u) \, = \, a_0 \,
  (\Delta + 1) \, u^4, \ \ \ \gamma (u) \, = \, - a_0 \, \frac{\Delta
    + 2}{2} \, u^4, \hspace*{1cm} \mbox{as} \hspace*{1cm} u
  \rightarrow 0.
\end{align}

Our goal now is to use the ansatz of Eqs. (\ref{aa}) and (\ref{bc})
and solve Einstein equations (\ref{eom}) in the $\tau \rightarrow 0$
limit keeping $u$ fixed. In this limit we write
\begin{subequations}\label{ABCans}
\begin{align}
  A (\tau, u) \, = \, e^{a (\tau, u)} \, = \, e^{\tau^{\Delta + 4} \,
    \alpha (u)} \, = \, 1 + \tau^{\Delta + 4} \, \alpha (u) \, + o
  (\tau^{2 \, \Delta + 8})
\end{align}
\begin{align}
  B (\tau, u) \, = \, 1 + \tau^{\Delta + 4} \, \beta (u) \, + o
  (\tau^{2 \, \Delta + 8})
\end{align}
\begin{align}
  C (\tau, u) \, = \, 1 + \tau^{\Delta + 4} \, \gamma (u) \, + o
  (\tau^{2 \, \Delta + 8})
\end{align}
\end{subequations}
as, with the ansatz of Eqs. (\ref{aa}) and (\ref{bc}), we do not have
any control over $o (\tau^{2 \, \Delta + 8})$ terms in $A$, $B$ and
$C$. 

Rewriting the metric (\ref{met_gen}) in terms of $\tau$, $u$, $\eta$,
and ${\un x} = (x_1, x_2)$ we plug it into \eq{eom}. At the lowest
order in $\tau$ we obtain the following equations, corresponding to
the $uu$, $u \, \tau$ and $x_1 x_1$ (or, equivalently, $x_2 x_2$)
components of the Einstein equations:
\begin{align}\label{uu}
  \frac{\alpha'}{2 \, u} + \frac{\beta'}{2 \, u} + \frac{\gamma'}{u} -
  \frac{\alpha''}{2} - \frac{\beta''}{2} - \gamma'' = 0,
\end{align}
\begin{align}\label{ut}
  \alpha' - \frac{3 + \Delta}{2} \, \beta' - (2 + \Delta) \, \gamma' -
\frac{1}{2} \, u \, \alpha'' = 0,
\end{align}
\begin{align}\label{xx}
  \left( 8 + 4 \, \Delta + \frac{1}{2} \, \Delta^2 \right) \, \gamma +
  \frac{\alpha'}{2 \, u} + \frac{\beta'}{2 \, u} + \left( \frac{5}{2
      \, u} - \frac{7}{2} \, u - \Delta \, u \right) \, \gamma' +
  \frac{1}{2} \, (u^2 -1) \, \gamma'' = 0.
\end{align}
Here $\alpha' = d \alpha (u) /du$, $\alpha'' = d^2 \alpha (u) /d u^2$,
and the same applies to derivatives of $\beta (u)$ and $\gamma (u)$.

After an integration using the initial conditions (\ref{init}), \eq{uu} gives
\begin{align}\label{traceless}
  \alpha + \beta + 2 \, \gamma =0. 
\end{align}
Using \eq{traceless} we eliminate $\gamma$ from \eq{ut} to obtain
\begin{align}\label{betaalpha}
  \beta \, = \, (5 + \Delta) \, \alpha - u \, \alpha'.
\end{align}
Substituting Eqs. (\ref{betaalpha}) and (\ref{traceless}) into \eq{xx}
yields
\begin{align}\label{aeq}
  & u^2 \, (u^2 -1) \, \alpha''' + u \left[ 7 + \Delta - (11 + 3 \,
    \Delta) \, u^2 \right] \, \alpha'' \notag \\ & + \left[ - 3 \, (5
    + \Delta) + ( 51 + 25 \, \Delta + 3 \, \Delta^2) \, u^2 \right] \,
  \alpha' - (4 + \Delta)^2 \, (6 + \Delta) \, u \, \alpha = 0.
\end{align}
The solution of \eq{aeq} satisfying the initial condition (\ref{init})
can be easily found by searching for it in the form of an infinite
series in $u$ starting at the order $u^4$. From \eq{ascal} we also
infer that the coefficient in front of the $u^5$ term in the series is
zero. The solution of \eq{aeq} satisfying such conditions is
\begin{align}
  \alpha (u) \, = \, a_0 \, u^4 \, F \left( -1 -
    \frac{\Delta}{2} , - \frac{\Delta}{2}; 3; u^2 \right) 
\end{align}
with $F$ the hypergeometric function. Eqs. (\ref{betaalpha}) and
(\ref{traceless}) allow us to find $\beta (u)$ and $\gamma (u)$. In
the end, using Eqs. (\ref{ABCans}), we obtain the following solution
for the components of the metric (\ref{met_gen})
\begin{subequations}\label{power_sol}
\begin{align}
  A (\tau, u) \,  = \, 1 + a_0 \, \tau^{4 + \Delta} \, u^4 & \, F \left(
    -1 - \frac{\Delta}{2} , - \frac{\Delta}{2}; 3; u^2 \right), \\
  B (\tau, u) \, = \, 1 + a_0 \, \tau^{4 + \Delta} \, u^4 & \, \bigg[
  (\Delta + 1) \,  F \left( -1 - \frac{\Delta}{2} , -
    \frac{\Delta}{2}; 3; u^2 \right) \notag \\  & - \frac{\Delta \,
    (\Delta + 2)}{6} \, u^2 \, F \left( 1 - \frac{\Delta}{2} , -
    \frac{\Delta}{2}; 4; u^2 \right) \bigg], \\
  C (\tau, u) \, = \, 1 + a_0 \, \tau^{4 + \Delta} \, u^4 & \ 
  \frac{\Delta + 2}{12} \, \bigg[ - 6 \,  F \left( -1 -
    \frac{\Delta}{2} , - \frac{\Delta}{2}; 3; u^2 \right) \notag \\ &
  + \Delta \, u^2 \, F \left( 1 - \frac{\Delta}{2} , -
    \frac{\Delta}{2}; 4; u^2 \right) \bigg].
\end{align}
\end{subequations}

As is well-known, hypergeometric functions, such as we have in
\eq{power_sol}, have a branch-cut discontinuity along the real axis
for $u \ge 1$.  Therefore, for the metric to be a real and
single-valued function of $u$ at $u > 1$ we need the hypergeometric
series to terminate. The reasoning here is the same as what we used in
re-deriving Bjorken hydrodynamics in Sect. \ref{Bjorken}.  As was
pointed out at the end of Sect. \ref{iter}, the allowed values of the
power $\Delta$ at early times are $-1 \le \Delta \le 0$. Within this
range it is only possible for the hypergeometric series to terminate
if $\Delta =0$. Therefore, $\Delta = 0$ is the only allowed solution!

Plugging in $\Delta = 0$ into \eq{power_sol} gives
\begin{subequations}\label{power_sol_fin}
\begin{align}
  A (\tau, u) \, &= \, 1 + a_0 \, \tau^{4} \, u^4 \\
  B (\tau, u) \, & = \, 1 + a_0 \, \tau^{4} \, u^4 \\
  C (\tau, u) \, & = \, 1 - a_0 \, \tau^{4} \, u^4.
\end{align}
\end{subequations}
Using $\Delta = 0$ in Eqs. (\ref{abc00}) and (\ref{epp3}) yields
\begin{align}\label{early1}
  \epsilon \, = \, - \frac{N_c^2}{2 \, \pi^2} \ a_0, \ \ \ p_3 \, = \,
  \frac{N_c^2}{2 \, \pi^2} \ a_0, \ \ \ p \, = \, - \frac{N_c^2}{2 \,
    \pi^2} \ a_0.
\end{align}
Note that at early times $a_0$ is different from the one used at
late times in Sect. \ref{Bjorken}. Still $a_0 < 0$ and we conclude
from \eq{early1} that, for the strongly-coupled medium described by
gauge-gravity duality in the geometry of heavy ion collisions
\begin{align}\label{earlye}
  \epsilon (\tau) \, \rightarrow \, \text{constant} \hspace*{1cm}
  \text{as} \hspace*{1cm} \tau \rightarrow 0.
\end{align}
We also obtain that, as $\tau \rightarrow 0$
\begin{align}\label{earlypp3}
  p_3 (\tau) \, = \, - \epsilon (\tau), \hspace*{1cm} p
  (\tau) \, = \, \epsilon (\tau).
\end{align}
The latter relations one can also obtain from Eqs. (\ref{hydroeq}) and
(\ref{trless}) by using \eq{earlye} in them. 

Note that for the description of heavy ion collisions in the framework
of the Color Glass Condensate (CGC), in which the produced medium has
a small coupling constant, it was suggested recently by Lappi
\cite{Lappi:2003bi} (see also \cite{Fukushima:2007ja}) that the
leading behavior of energy density as $\tau \rightarrow 0$ is given by
\begin{align}\label{ecgc}
  \epsilon_{\text{CGC}} (\tau) \, \propto \, \ln^2 \left(
    \frac{1}{\tau} \right) \hspace*{1cm} \text{as} \hspace*{1cm} \tau
  \rightarrow 0.
\end{align}
Using \eq{ecgc} in Eqs. (\ref{hydroeq}) and (\ref{trless}) and keeping
leading logarithms only one arrives at \eq{earlypp3}. Hence, while
\eq{ecgc} is indeed different from our \eq{earlye}, the energy density
and pressures at early proper times behave somewhat similar for the
cases of CGC and the strongly coupled medium considered above. In both
cases the initial longitudinal pressure is negative, while the
transverse pressure is positive and equal to the energy density.

%%%%%%%%%%%%%%%%%%%%%%%%%%%%%%%%%%%%%%%%%%%%%%%%%%%%%%%%%%%%%%%%%%%%%%%%%%%%%%%%%

\section{Logarithmic scaling of energy density \\ with proper time}
\label{log}

In the previous Section we looked for the energy density at early
proper times using the power-law ansatz (\ref{etau}) and found that
the power of $\tau$ has to be zero. However, in view of the CGC result
(\ref{ecgc}) \cite{Lappi:2003bi} it may be that the absence of a
power-law scaling indicates some residual logarithmic scaling of
energy density with proper time. Here we will verify whether such
logarithmic scaling does indeed take place.

It is tempting to begin by assuming that the energy density scales
logarithmically with $\tau$ such that
\begin{align}\label{elog}
  \epsilon (\tau) \, \propto \, \ln^\delta \left( \frac{1}{\tau}
  \right)
\end{align}
with some number $\delta \ge 0$. However, if we try solving Einstein
equations (\ref{eom}) with the ansatz (\ref{elog}) at $z=0$ keeping
the leading logarithmic [$\ln (1/\tau)$] terms at small $\tau$ we will
simply reproduce the results of the previous Section [with the second
terms in each of the equations (\ref{power_sol_fin}) multiplied by
$\ln^\delta \left( \frac{1}{\tau} \right)$] without obtaining any
constraints on $\delta$. The reason for that is quite straightforward:
if one differentiates a power of $\ln (1/\tau)$ with respect to $\tau$
one power of the logarithm would be lost and such contribution would
not be leading logarithmic anymore and would have to be discarded.
Therefore, in the leading logarithmic approximation no derivative acts
on the power of the logarithm, making $ \ln^\delta \left(
  \frac{1}{\tau} \right)$ term just an overall factor canceling out in
the Einstein equations and not giving any constraint on $\delta$.

To avoid this problem one has to include the subleading logarithmic
correction. We start by writing
\begin{align}\label{a0log}
  a_0 (\tau) \, = \, a_0 \, \ln^\delta \left( \frac{1}{\tau} \right)
  + a_1 \, \ln^{\delta -1} \left( \frac{1}{\tau} \right) + \ldots
\end{align}
such that, using \eq{epp3},
\begin{align}\label{elog2}
  \epsilon (\tau) \, = \, - \frac{N_c^2}{2 \, \pi^2} \, \left[ a_0 \,
    \ln^\delta \left( \frac{1}{\tau} \right) + a_1 \, \ln^{\delta
      -1} \left( \frac{1}{\tau} \right) + \ldots \right].
\end{align}
Here $a_1$ is some undetermined constant. The ellipsis denote terms of
the order $o \left[ \ln^{\delta -2} \left( \frac{1}{\tau} \right)
\right]$. Using \eq{elog2} in Eqs.  (\ref{hydroeq}) and (\ref{trless})
yields
\begin{subequations}\label{b0c0log}
\begin{align}
  b_0 (\tau) \, = \, a_0 \, \ln^\delta \left( \frac{1}{\tau} \right)
  + (a_1 - a_0 \, \delta) \, \ln^{\delta -1} \left( \frac{1}{\tau} \right) + \ldots \\
  c_0 (\tau) \, = \, - a_0 \, \ln^\delta \left( \frac{1}{\tau} \right)
  + \left( \frac{a_0 \, \delta}{2} - a_1 \right) \, \ln^{\delta -1}
  \left( \frac{1}{\tau} \right) + \ldots.
\end{align}
\end{subequations}

We performed a perturbative expansion of the solution of Einstein
equations, similar to the one carried out in Sect.  \ref{iter}. It
indicated that the scaling variable is still $u$ defined in \eq{u}.
Therefore we try looking for the solution of \eq{eom} in the following
form
\begin{subequations}\label{abclog}
\begin{align}
  A (\tau, u) \, &= \, 1 + a_0 \, \tau^{4} \, u^4 \, \ln^\delta \left(
    \frac{1}{\tau} \right) + \tau^{4} \, \alpha_1 (u) \,
  \ln^{\delta -1} \left( \frac{1}{\tau} \right)  + \ldots \\
  B (\tau, u) \, & = \, 1 + a_0 \, \tau^{4} \, u^4 \, \ln^\delta
  \left( \frac{1}{\tau} \right) + \tau^{4} \, \beta_1 (u) \,
  \ln^{\delta -1} \left( \frac{1}{\tau} \right)  + \ldots \\
  C (\tau, u) \, & = \, 1 - a_0 \, \tau^{4} \, u^4 \, \ln^\delta
  \left( \frac{1}{\tau} \right) + \tau^{4} \, \gamma_1 (u) \,
  \ln^{\delta -1} \left( \frac{1}{\tau} \right) + \ldots
\end{align}
\end{subequations}
with the unknown functions $\alpha_1 (u)$, $\beta_1 (u)$ and $\gamma_1
(u)$ satisfying the initial conditions which follow from Eqs.  (\ref{a0log}),
(\ref{b0c0log})
\begin{align}\label{abg1}
  \alpha_1 (u) \, = \, a_1 \, u^4, \ \ \ \beta_1 (u) \, = \, (a_1 -
  a_0 \, \delta) \, u^4, \ \ \ \gamma_1 (u) \, = \, \left( \frac{a_0
      \, \delta}{2} - a_1 \right) \, u^4, \hspace*{1cm} \mbox{as}
  \hspace*{1cm} u \rightarrow 0.
\end{align}

The ansatz of \eq{abclog} trivially satisfies the Einstein equations
at the leading logarithmic order for the reasons discussed above.
Non-trivial information comes in at the subleading logarithmic order.
At the order $\ln^{\delta -1} \left( \frac{1}{\tau} \right)$ the $uu$,
$x_1 x_1$ and $u x_1$ components of the Einstein equations (\ref{eom})
read
\begin{align}\label{uulog}
  \frac{\alpha'_1}{2 \, u} + \frac{\beta'_1}{2 \, u} +
  \frac{\gamma'_1}{u} - \frac{\alpha''_1}{2} - \frac{\beta''_1}{2} -
  \gamma''_1 \, = \, 0,
\end{align}
\begin{align}\label{xxlog}
  16 \, u \, \gamma_1 + \alpha'_1 + \beta'_1 + (5 - 7 \, u^2) \,
  \gamma'_1 + u \, (u^2 -1) \, \gamma''_1 \, = \, 0,
\end{align}
\begin{align}\label{xulog}
  4 \, a_0 \, \delta \, u^3 - 2 \, \alpha'_1 + 3 \, \beta'_1 + 4 \,
  \gamma'_1 + u \, \alpha''_1 \, = \, 0.
\end{align}
Just like in Sect. \ref{earlytime} we integrate \eq{uulog} to obtain
\begin{align}\label{tracelesslog}
  \alpha_1 + \beta_1 + 2 \, \gamma_1 \,  = \, 0. 
\end{align}
Using \eq{tracelesslog} in \eq{xxlog} to eliminate $\alpha'_1 +
\beta'_1$ yields
\begin{align}\label{g1}
  16 \, u \, \gamma_1 + (3 - 7 \, u^2) \, \gamma'_1 + u \, (u^2 -1) \,
  \gamma''_1 \, = \, 0.
\end{align}
The only solution of \eq{g1} satisfying the initial conditions
(\ref{abg1}) is
\begin{align}\label{gamma1log}
  \gamma_1 (u) \, = \, \left( \frac{a_0 \, \delta}{2} - a_1 \right)
  \, u^4. 
\end{align}
Using the solution (\ref{gamma1log}) in Eqs. (\ref{xulog}) and
(\ref{tracelesslog}) yields
\begin{align}\label{a1log}
  u \, \alpha''_1 - 5 \, \alpha'_1 + 8 \, a_1 \, u^3 \, = \, 0.
\end{align}
The solution of \eq{a1log} satisfying the initial conditions
(\ref{abg1}) is
\begin{align}\label{alpha1log}
  \alpha_1 (u) \, = \, a_1 \, u^4.
\end{align}
Finally, plugging Eqs. (\ref{gamma1log}) and (\ref{alpha1log}) into
\eq{tracelesslog} yields
\begin{align}\label{beta1log}
  \beta_1 (u) \, = \, (a_1 - a_0 \, \delta) \, u^4.
\end{align}
To make sure that Eqs. (\ref{alpha1log}), (\ref{beta1log}) and
(\ref{gamma1log}) give us the solution of \eq{eom} we need to verify
that the remaining two equations corresponding to $\eta \eta$ and
$\tau \tau$ components of \eq{eom} are satisfied. These read
\begin{align}\label{eelog}
  2 \, a_0 \, \delta \, u^5 - 4 \, u \, \alpha_1 + 20 \, u \, \beta_1
  + 8 \, u \, \gamma_1 + (1 + u^2) \, \alpha'_1 + 4 \, (1 - 2 \, u^2)
  \, \beta'_1 + 2 \, (1 - u^2) \, \gamma'_1 \notag \\ - u \, (1 - u^2)
  \, \beta''_1 \, = \, 0
\end{align}
and
\begin{align}\label{ttlog}
  - 6 \, a_0 \, \delta \, u^5 + 4 \, u \, \alpha_1 - 20 \, u \,
  \beta_1 - 24 \, u \, \gamma_1 + 2 \, (u^2 - 2) \, \alpha'_1 - (u^2 -
  1) \, (- \beta'_1 - 2 \, \gamma'_1 + u \, \alpha''_1) \, = \, 0.
\end{align}
Substituting Eqs. (\ref{alpha1log}), (\ref{beta1log}) and
(\ref{gamma1log}) into Eqs. (\ref{eelog}) and (\ref{ttlog}) we see
that the latter equations can be satisfied for $a_0 \neq 0$ only if
\begin{align}
  \delta \, = \, 0.
\end{align}
This constrains the power of the logarithm in \eq{elog}. Putting
$\delta =0$ in \eq{elog2} we obtain
\begin{align}\label{elog3}
  \epsilon (\tau) \, = \, - \frac{N_c^2}{2 \, \pi^2} \, \left[ a_0 +
    \frac{a_1}{\ln (1/\tau)} + \ldots \right], \hspace*{1cm} \text{as}
  \hspace*{1cm} \tau \rightarrow 0.
\end{align}
We therefore confirm the result of the previous Section that the
energy density approaches a constant as $\tau \rightarrow 0$.
\eq{elog3} demonstrates that the approach to the constant may be
logarithmic.

Again our derivation never employed the requirement of no
singularities of the curvature invariant ${\cal R} =
R_{\mu\nu\rho\sigma} \, R^{\mu\nu\rho\sigma}$ of \cite{Janik:2005zt}.
At the same time it can be explicitly checked that it does not violate
such a requirement. The curvature invariant for the solution in
\eq{power_sol_fin} (which is also the leading small-$\tau$ solution
for the logarithmic scaling considered in this Section, as follows
from the above) is
\begin{align}\label{RR}
{\cal R} \, = \, 40 + o (\tau^8).
\end{align}
Here we keep $u$ fixed and take $\tau$ to be small. \eq{RR} shows that
the non-trivial part of the solution (\ref{power_sol_fin}) does not
contribute to the curvature invariant and, therefore, does not
introduce singularities in it.  In fact one can show in general that
the first perturbative correction to the metric of the flat AdS space
satisfying \eq{eom} does not contribute to the curvature invariant. It
appears that it would have been impossible to constrain the early
times scaling of the energy density using non-singularity of the
curvature invariant only. Luckily, we were able to pinpoint the right
metric without imposing any additional conditions.

%%%%%%%%%%%%%%%%%%%%%%%%%%%%%%%%%%%%%%%%%%%%%%%%%%%%%%%%%%%%%%%%%%%%%%%%%%%%%%%%%

\section{Estimate of isotropization time}
\label{iso_time}

We have shown above that the strongly-coupled system produced in heavy
ion collisions with the dynamics described by the AdS/CFT
correspondence starts off at early times with a finite energy density
and negative longitudinal pressure component of the energy-momentum
tensor.  Hence, the system at these early times is very asymmetric. On
the other hand, as was demonstrated by Janik and Peschanski
\cite{Janik:2005zt}, the same strongly-coupled system at
asymptotically late times behaves like the ideal Bjorken
hydrodynamics, with the longitudinal and transverse pressure
components equal and positive. Therefore, the dynamics of this
strongly-coupled system incorporates the isotropization transition at
some intermediate time, and has the potential of solving the
thermalization problem in heavy ion collisions. Indeed, since the
initial condition in real heavy ion collisions described by QCD are
likely to be perturbative describable by CGC, to verify the relevance
of the strongly-coupled model considered above to the real-life heavy
ion collisions it is important to check whether the early times
dynamics obtained here agrees with perturbative expectations. The very
early time dynamics was studied in the CGC framework in
\cite{Lappi:2003bi} with the conclusion that the energy density scales
as shown in \eq{ecgc}. Such behavior also leads to negative initial
longitudinal pressure, similar to what we observed for the
strongly-coupled theory.  Therefore it is plausible that the
strongly-coupled isotropization scenario is relevant for the real-life
heavy ion collisions described by QCD.

With that in mind let us estimate the onset of isotropic Bjorken
hydrodynamics in the gauge-gravity duality framework considered here.
First of all, at early times the general solution (\ref{power_sol})
has a branch-cut discontinuity at $u \ge 1$. In the $z, \tau$-space
the branch cut is at
\begin{align}\label{zearly}
z \, \ge \, z_c^\text{early} \, \equiv \, \tau. 
\end{align}
On the other hand, the solution found by Janik and Peschanski
\cite{Janik:2005zt} and shown here in \eq{JPsol} has a branch cut at
\begin{align}\label{vlate}
v \, \ge \, D^{-\frac{1}{4}}
\end{align}
which, with the help of \eq{v} with $\Delta = -4/3$ and corresponding
$D=1/3$ translates into
\begin{align}\label{zlate}
  z \, \ge \, z_c^\text{late} \, \equiv \, \tau^{\frac{1}{3}} \,
  \left( \frac{3}{-a_0} \right)^\frac{1}{4}.
\end{align}

\begin{figure}[h]
\begin{center}
\epsfxsize=10cm
\leavevmode
\hbox{\epsffile{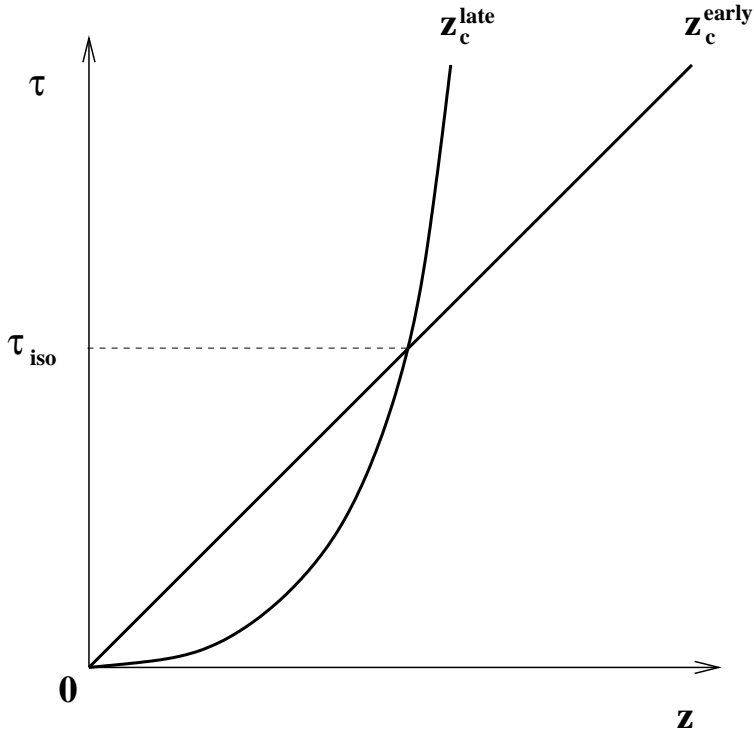}}
\end{center}
\caption{The locations of branching points $z_c^\text{early}$ and 
  $z_c^\text{late}$ for the early- and late-time metrics in the $z,
  \tau$-plane. The intersection of the two curves gives our estimate
  of the isotropization time $\tau_\text{iso}$ (see text). }
\label{iso}
\end{figure}

The locations of the starting points of the branch cuts
$z_c^\text{early}$ and $z_c^\text{late}$ given by Eqs. (\ref{zearly})
and (\ref{zlate}) are shown in the $z, \tau$-plane in \fig{iso}. The
two curves in \fig{iso} intercept at some time which we label
$\tau_\text{iso}$.

%%%%%%%%%%%%%%%%%%%%%%%%%%%%%
%%%%%%%%%%%%%%%%%%%%%%%%%%%%%%

It is important to point out that, as can be seen, say, from
\eq{abc2}, the full solution with the power-law ansatz of \eq{etau}
consists of a sum (for $a$, $b$ and $c$) of the solution in \eq{JPsol}
\cite{Janik:2005zt} and the solution in \eq{power_sol} found above.
While different solutions dominate at different times, both exist at
all times. We now observe that in order to pinpoint the correct
solution using the power-law ansatz of \eq{etau} as was done above one
starts outside the branch cut, i.e., at $z < z_c$ for both early and
late times, and imposes the no-branch-cut condition as $z$ increases
past $z_c$. For the times before the intersection of the two curves in
\fig{iso}, $\tau < \tau_\text{iso}$, the first branch cut which has to
be eliminated is the one starting at $z_c^\text{early}$. That fixes
the power $\Delta =0$ making the energy density a constant, as
described above and shown in \eq{earlye}.  At late times the first
branch cut to be eliminated is at $z_c^\text{late}$, fixing $\Delta =
-4/3$ as described in Section \ref{Bjorken}. Hence the behavior of
energy density, or, equivalently, the dominance of either one of the
solutions in Eqs. (\ref{JPsol}) and (\ref{power_sol}), is determined
by which one of the two branch cuts starts at smaller values of $z$.
At $z_c^\text{early} < z_c^\text{late}$ the early-time solution
dominates leading to the energy density scaling of \eq{earlye}. At
$z_c^\text{early} > z_c^\text{late}$ the late-time solution dominates
leading to Bjorken's energy density scaling of \eq{ebj}. While,
strictly speaking both solutions are valid only at either
asymptotically early or late times far away from the intersection
point shown in \fig{iso}, we can estimate the time of the transition
between the two regimes by the time of the intercept
$\tau_\text{iso}$, defined by the condition
\begin{align}\label{zz}
z_c^\text{early} \, = \, z_c^\text{late}.
\end{align}
Using Eqs. (\ref{zearly}) and (\ref{zlate}) in \eq{zz} yields
\begin{align}\label{tiso1}
\tau_\text{iso} \, = \, \left( \frac{3}{-a_0} \right)^\frac{3}{8}.
\end{align}
This is our estimate of the isotropization time, the time necessary
for hydrodynamics to work and required for thermalization. To
determine $a_0$ we write Bjorken energy density scaling as
\begin{align}\label{ebj1}
\epsilon (\tau) \, = \, \frac{\epsilon_0}{\tau^\frac{4}{3}}.
\end{align}
Comparing Eqs. (\ref{ebj1}) and (\ref{ebj}) we see that
\begin{align}
a_0 \, = \, - \frac{2 \, \pi^2}{N_c^2} \, \epsilon_0,
\end{align}
which, when substituted into \eq{tiso1} gives
\begin{align}\label{tiso}
  \tau_\text{iso} \, = \, \left( \frac{3}{\epsilon_0} \,
    \frac{N_c^2}{2 \, \pi^2} \right)^\frac{3}{8}.
\end{align}
Note that for a self-consistent conformally-invariant strongly
interacting theory the above result is natural: from the scaling of
\eq{ebj1} we see that such theory would be characterized only by one
dimensionful parameter --- $\epsilon_0$. Using this parameter we can
construct the time scale, which is $\epsilon_0^{-3/8}$. Indeed to find
the coefficient in front of $\epsilon_0^{-3/8}$ in \eq{tiso} one needs
to perform a more complete calculation, as was done above.

To evaluate the isotropization time in \eq{tiso} for central $Au+Au$
collisions with $\sqrt{s} = 200$~GeV at RHIC we use the fact that
hydrodynamic simulations of
\cite{Kolb:2000sd,Kolb:2000fh,Huovinen:2001cy,Kolb:2001qz,Heinz:2001xi},
which successfully describe a variety of RHIC observables, yield the
averaged over all impact parameters energy density of $\epsilon =
15$~GeV/fm$^3$ \cite{Kolb:2003dz,Heinz:2001xi} at the proper time
$\tau = 0.6$~fm/c.  From these data, using \eq{ebj1} we obtain
$\epsilon_0 \approx 38$~fm$^{-8/3}$.  Substituting this number into
\eq{tiso} with $N_c =3$ yields
\begin{align}\label{tiso_num}
  \tau_\text{iso} \, \approx \, 0.29 \, \text{fm/c}
\end{align}
which is quite close to the initialization time for the hydrodynamic
simulations required to describe RHIC data
\cite{Kolb:2000sd,Kolb:2000fh,Huovinen:2001cy,Kolb:2001qz,Heinz:2001xi,Kolb:2003dz,Teaney:1999gr,Teaney:2000cw,Teaney:2001av}!

Indeed the isotropization time estimate of \eq{tiso} depends on the
final state observable $\epsilon_0$ instead of some initial-state
observable. Therefore one should indeed question the predictive power
of such estimate. We leave it for the future work to improve on this
result and to tie it to some initial state observables.

%%%%%%%%%%%%%%%%%%%%%%%%%%%%%%%%%%%%%%%%%%%%%%%%%%%%%%%%%%%%%%%%%%%%%%%%%%%%%%%%%

\section{Conclusions}
\label{conc}

We conclude by restating our main results. We have re-derived the
result of Janik and Peschanski \cite{Janik:2005zt} stating that at
late proper times the system produced in heavy ion collisions in the
strongly-coupled regime described by gauge-gravity duality exhibits
energy density scaling with time characteristic of Bjorken
hydrodynamics, as shown in \eq{ebj}. In such regime all three pressure
components of the energy momentum tensor are equal to each other, as
is required for a thermalized medium. To obtain this result we use a
much weaker condition than that used in \cite{Janik:2005zt}: we
require the metric to be a real and single-valued function of
coordinates in the bulk.

We then explore the dynamics of the same strongly-coupled system at
early times and show that the energy density approaches a constant as
$\tau \rightarrow 0$, as shown in \eq{earlye}. The longitudinal
pressure of this early system is therefore negative [see
\eq{hydroeq}]. We have therefore demonstrated that this
strongly-coupled system is initially anisotropic, but at late times it
becomes isotropic and evolves according to Bjorken hydrodynamics.
Therefore, a complete solution for the energy density of this system
at all times would contain the isotropization transition! The
similarity of the early-time behavior of the system to that of the CGC
demonstrates that our conclusions may be relevant to heavy ion
collisions in the real world.

%%%%%%%%%%%%%%%%%%%%%%%%%%%%%
\begin{figure}[h]
\begin{center}
\epsfxsize=10cm
\leavevmode
\hbox{\epsffile{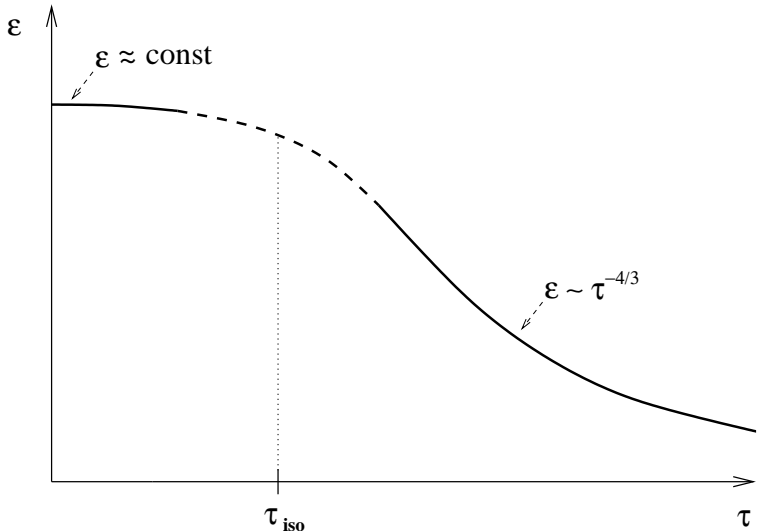}}
\end{center}
\caption{The scaling of energy density $\epsilon$ with proper time 
  $\tau$ in heavy ion collisions for a strongly-coupled system as
  described by AdS/CFT correspondence. At late time Bjorken
  hydrodynamics is recovered, as was shown in \cite{Janik:2005zt}. At
  early times the energy density goes to a constant, as shown in this
  work. Our estimate of the isotropization time $\tau_\text{iso}$ is
  likely located in some transition (isotropization) region. }
\label{escaling}
\end{figure}
%%%%%%%%%%%%%%%%%%%%%%%%%%%%%%

\fig{escaling} summarizes our knowledge of the energy density in heavy
ion collisions as a function of $\tau$ for the strongly-coupled system
considered above and in \cite{Janik:2005zt}. The energy density starts
out as a constant, after which the system undergoes an isotropization
transition (denoted by the dashed line in \fig{escaling}) resulting in
Bjorken hydrodynamics. The exact shape of the dashed curve is indeed
unknown: what we plot in \fig{escaling} is a simple interpolation
between the early- and late-time regimes. More work is needed to
develop a better understanding of this important intermediate region.

Finally, above we estimated the time of the transition from our regime
to that of Janik and Peschanski. The estimate is given in \eq{tiso}.
If applied to RHIC, it appears to give the isotropization time
$\tau_\text{iso} \, \approx \, 0.3$~fm/c, consistent with the results
of hydrodynamic simulations
\cite{Kolb:2000sd,Kolb:2000fh,Huovinen:2001cy,Kolb:2001qz,Heinz:2001xi,Teaney:1999gr,Teaney:2000cw,Teaney:2001av}.

In conclusion let us point out that the method of requiring the metric
to be a real and single-valued function of coordinates could be
successfully applied to other related problems. Recently
$o(1/\tau^{2/3})$ and $o(1/\tau^{4/3})$ corrections to the solution of
\eq{JPsol} have been calculated in \cite{Nakamura:2006ih,Janik:2006ft}
for late proper times. One expands \cite{Nakamura:2006ih,Janik:2006ft}
\begin{align}\label{aaa}
  a (\tau, z) \, = \, a(v) + a_1 (v) \, \frac{1}{\tau^{2/3}} + a_2 (v)
  \, \frac{1}{\tau^{4/3}} + \ldots
\end{align}
and finds the subleading coefficients $a_1 (v)$ and $a_2 (v)$ by
solving Einstein equations perturbatively. [The same expansion is done
for $b(\tau,z)$ and $c (\tau,z)$.] Taking the $o(1/\tau^{4/3})$
correction to the coefficient $a(v)$ in \eq{JPsol}, denoted by $a_2
(v)$ and given in Eq. (20) of \cite{Janik:2006ft}, one can see that it
also has a term proportional to $\ln (1 - v^4 /3)$, just like the
solution in \eq{JPsol} with $\Delta = -4/3$:
\begin{align}\label{a2}
  a_2 (v) \, = \, \left( \frac{1}{4 \, \sqrt{3}} - \frac{3 \,
      \eta_0^2}{2} \right) \, \ln (1 - D \, v^4) + \text{terms without
    branch cuts}.
\end{align}
Here $\eta_0$ is the proper-time independent coefficient related to
shear viscosity $\eta$ by \cite{Nakamura:2006ih,Janik:2006ft}
\begin{align}
  \eta \, = \, \frac{\eta_0}{\tau}.
\end{align}
Again we want the metric to be real and single-valued, hence the
coefficient in front of $\ln (1 - D \, v^4)$ has to be an integer.
Note however that $a_2 (v)$ enters the metric with a prefactor of
$1/\tau^{4/3}$ as can be seen from \eq{aaa}, such that the coefficient
in front of the logarithm in the metric varies with time. Therefore we
conclude that the coefficient in front of $\ln (1 - D \, v^4)$ can
only be zero: this would insure that the metric is real at all times.
Putting the coefficient in front of $\ln (1 - D \, v^4)$ in \eq{a2} to
zero yields
\begin{align}
  \eta_0^2 \, = \, \frac{\sqrt{3}}{18}
\end{align}
in agreement with the result obtained in \cite{Janik:2006ft} using
non-singularity of the curvature invariant criterion, and, as shown in
\cite{Janik:2006ft}, in agreement with the
Kovtun-Policastro-Son-Starinets (KPSS) viscosity bound
\cite{Policastro:2001yc,Son:2002sd,Policastro:2002se,Kovtun:2003wp,Kovtun:2004de}!

Here we have to caution the reader that similar branch cuts in other
metric coefficients, like $b_1 (v)$, $b_2 (v)$ and $c_1 (v)$, $c_2
(v)$ of \cite{Janik:2006ft} can not be eliminated this easily. ($a_1
(v)$ has no branch cut singularity \cite{Janik:2006ft}.) It is
possible that resummation of the whole series of the type shown in
\eq{aaa} for $b (\tau, z)$ and $c (\tau, z)$ would be required to
eliminate such branch cuts. We leave the investigation of this
important question for future work.

%%%%%%%%%%%%%%%%%%%%%%%%%%%%%%%%%%%%%%%%%%%%%%%%%%%%%%%%%%%%%%%%%%%%%%%%%%%%%%%

\section*{Acknowledgments} 

We would like to thank Ulrich Heinz, Romuald Janik, Robi Peschanski,
Dam Son and Heribert Weigert for stimulating and informative
discussions.

This research is sponsored in part by the U.S. Department of Energy
under Grant No. DE-FG02-05ER41377.

%%%%%%%%%%%%%%%%%%%%%%%%%%%%%%%%%%%%%%%%%%%%%%%%%%%%%%%%%%%%%%%%%%%%%%%%%%%%%%

%\bibliography{references}                   
%\bibliographystyle{JHEP}

%%%%%%%%%%%%%%%%%%%%%%%%%%%%%%%%%%%%%%%%%%%%%%%%%%%%%%%%%%%%%%%%%%%%%%%%%%%%%%

\providecommand{\href}[2]{#2}\begingroup\raggedright\endgroup

\end{document}